# A SOLUTION FOR ARP SPOOFING: LAYER-2 MAC AND PROTOCOL FILTERING AND ARPSERVER

Yuksel Arslan

**ABSTRACT**

Most attacks are launched inside the companies by the employees of the same company. These kinds of attacks are generally against layer-2, not against layer-3 or IP. These attacks abuse the switch operation at layer-2. One of the attacks of this kind is Address Resolution Protocol (ARP) Spoofing (sometimes it is called ARP poisoning). This attack is classified as the "man in the middle" (MITM) attack. The usual security systems such as (personal) firewalls or virus protection software can not recognize this type of attack. Taping into the communication between two hosts one can access the confidential data. Malicious software to run internal attacks on a network is freely available on the Internet, such as Ettercap.

In this paper a solution is proposed and implemented to prevent ARP Spoofing. In this proposal access control lists (ACL) for layer-2 Media Access Control (MAC) address and protocol filtering and an application called ARPserver which will reply all ARP requests are used.

## 1. INTRODUCTION

Nowadays Ethernet is the most common protocol used at layer-2 of Local Area Networks (LANs). Ethernet protocol is implemented on the Network Interface Card (NIC). On top of Ethernet, Internet Protocol (IP), Transmission Control/User Datagram Protocols (TCP/UDP) are employed respectively. In this protocol stack for a packet to reach its destination IP and MAC of destination have to be known by the source. This can be done by ARP which is a protocol running at layer-3 of Open System Interface (OSI) model. ARP is designed without considering security issues like all other TCP/IP protocol stack. Malicious people can use these protocols for their own use.

ARP [1] finds the MAC address of destination computer by using the IP address of destination computer. ARP spoofing forces the destination computer to send packets to the attacker instead of source. Attacker can tap into the communication by forcing source and destination computers to send packets to itself at the same time. When it comes to this situation it is called MITM. In this situation the attacker can read packets of source and destination computers, save them and may change packet content or inject new packets. Figure 1 is showing MITM.

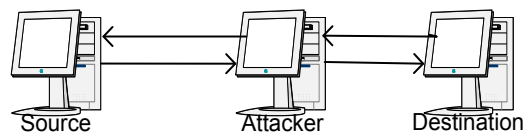

**Figure 1 MITM**

This paper is organized as follows: Section 2 describes details of ARP and ARP spoofing. Section 3 describes previous work on the subject and other protection methods. The proposal and its implementation are explained in Section 4. Section 5 concludes the paper.

## 2. PROBLEM DESCRIPTION
### 2.1 ARP
ARP is a protocol used in the TCP/IP protocol suite at the internet layer. TCP/IP protocol suite is shown in Figure 2 below.

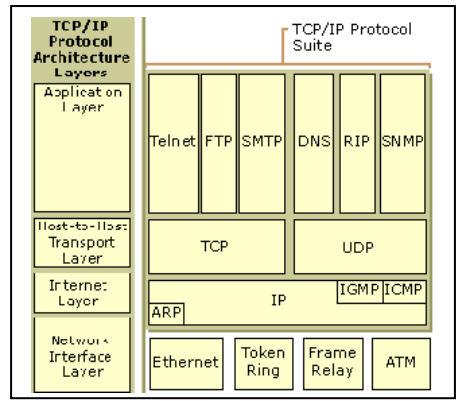

**Figure 2 TCP/IP Protocol Stack**

At Ethernet LANs using TCP/IP, address of a computer is determined by <IP, MAC> pair. IP is the 32 bit IPv4 address which is added to packet header. This address is assigned by the network administrator, Dynamic Host Configuration Protocol (DHCP) or owner of the computer as unique in that LAN. Of course if it is directly connected to internet it should be assigned by Internet Service Provider. IP is unique in that LAN.

MAC address is 48 bit LAN interface address of a NIC. Every NIC has a unique MAC address in the World. ARP provides mapping between IP and MAC address. In other words ARP finds the MAC address of the destination computer by using the IP address of the destination computer. If a computer has a packet to send, first it checks the ARP cache. ARP cache is the place where IP and MAC address mappings (<IP, MAC> pairs) are kept for a limited time. ARP cache mappings have aging time. At the end of this aging time <IP, MAC> pairings are flushed from the cache. This is because of dynamic nature of LANs. A new computer can be added or a computer may be shutdown, so it is meaningful that these cache entries have a limited life. If source computer can not find the MAC address of the destination computer, it broadcasts an ARP request packet. Every host on the network receives the request and checks if the IP address in the request is bound to one of its network interfaces. If this is the case, host with the matching IP address sends a unicast reply to the sender of request with its <IP, MAC> pair. This <IP, MAC> pair is entered into the ARP cache for further use. Next time when this host wants to send a packet to the same destination it retrieves the MAC address from the cache thus reducing the broadcast packet count. Figure 3 shows the ARP request/reply packet content.

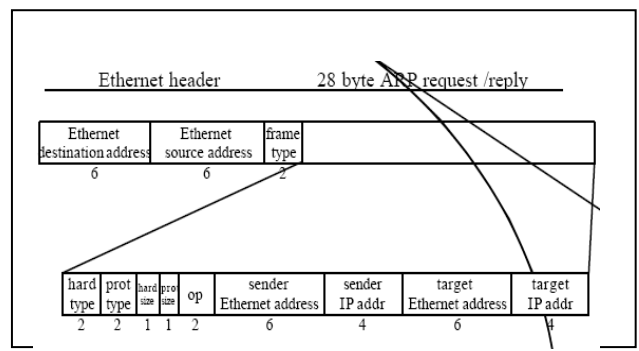

**Figure 3 ARP request/replay packet [2]**

ARP is stateless protocol, i.e., a reply may be processed even though the corresponding request was never sent. When a host receives a reply, it updates the corresponding entry in the cache with the <IP, MAC> pair in the reply. All possible cases that result cache update as follows:

**Solicited entries:** A host sends an ARP request and gets the reply. The IP and the MAC of this reply are entered into the cache. These are solicited entries, after a lifetime they are deleted from the cache.

**Unsolicited entries:** A host can receive an ARP request that is not destined for it, because ARP packets are broadcast. Even though it is not destined for itself, it checks if the sender IP is in the cache, if it is in the cache then it updates MAC of this IP with the new one received. Some ARP implementations may enter this <IP, MAC> pair if not present in the cache as a new entry as contrary to the RFC 826 [1].

**Static entries:** These entries are entered manually and they have no lifetime. They live until next reboot. They are not updated by ARP request or reply packets.

**Gratuitous ARP:** A gratuitous ARP is a message sent by a host requesting the MAC address for its own IP address. It is sent either by a host that wishes to determine if there is another host on the LAN with the same IP address or by a host announcing that it has changed its MAC address, thus allowing the other hosts to update their caches [3]. Figure 4 shows how ARP works.

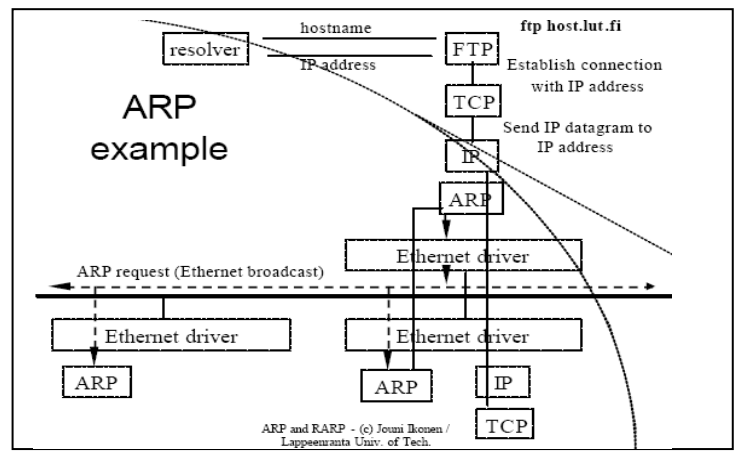

**Figure 4 ARP operation [2]**

## 2.2 ARP Spoofing

By forging an ARP reply, an attacker may easily change the <IP, MAC> association contained in a host ARP cache. Since each host presumes its local cache to be trustworthy the spoofed host will send IP packets encapsulated into Ethernet frames with a bogus MAC address as destination. This way the attacker may receive all the frames originally directed to some other host. If also the cache of the real destination host is poisoned, both communication flows are under the attacker's control.

Figure 5 shows the caches of the spoofed source and destination hosts.

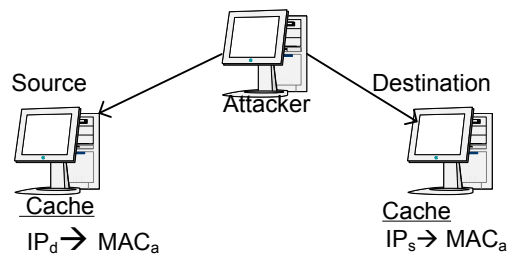

**Figure 5 ARP cache of source and destination hosts after attack**

The attacker realizes a two-way man in the middle, where he/she can forward the received packets to the correct destination after inspecting and possibly modifying them. The two end points of the connection will not notice the extra hop added by the attacker if the packet TTL is not decremented.

Some operating systems, will not update an entry in the cache if such an entry is not already present when an unsolicited ARP reply is received. The attacker can easily circumvent this precaution as follows: The attacker needs to trick the victim into adding a new entry in the cache first, so that a future ARP reply can update it. By sending a forged Internet Message Control Protocol (ICMP) echo request, it replies with an ICMP echo reply, which requires resolving first the IP address of the original ICMP request into an MAC address, thus creating an entry in the cache. The attacker can now update it with an unsolicited ARP reply [3].

## 3. RELATED WORK AND OTHER PREVENTION METHODS

There are some studies made up to now to prevent ARP spoofing. In [4] several studies which were categorized as detection, mitigation, and prevention or blocking were investigated and their advantages and disadvantages were explained. Security companies are also trying to find a solution to this problem.

There are actually two ways of preventing ARP attacks. One is changing the ARP protocol itself, by adding security or completely changing the protocol implementation. Some of these are not backward compatible as in [5] and [6] and some are as in [3]. Because it is possible to change the ARP cache of a host by directly sending an ARP reply packet to that host, without other hosts' noticing, the only way to prevent ARP spoofing without changing ARP protocol is programming the switch. All layer-2 frames are flowed under the control of switches. Second way of resolving ARP attacks is to interact with the switch to get the required information to realize if an ARP packet is an attack or not. Solutions explained in [7] and [8] fall into this category. Additionally these can be monitored and controlled at a central palace. Solution explained in this paper is also in this category.

In the following paragraphs a few of the solutions will be explained in more detail.

### 3.1 ARP-Guard box [7]:

It is a solution of ISL and SECUDOS firms which consists of hardware and software. DOMOS operating system runs on the hardware and the ARP-Guard software runs on this operating system. The ARP-Guard software has also a version which runs on Linux and Windows. ARP-Guard is based on sensors. It uses two sensors, LAN and SNMP sensors. LAN sensor listens to the mirror port of the switch and alarms the management console about ARP spoofing attacks. SNMP sensor is used for to get ARP information from routers. After sensors alert the management console, management console inform network administrator or can take preventive action automatically.

When using ARP-Guard it should be noted that mirror ports may not catch all the packets all the time, it is possible to miss some ARP packets and as a consequence some attacks. SNMP just get the ARP information from router thus if the attacker and the victim are in the same subnet there is no attack detection [8].

### 3.2 S-ARP (Secure ARP) [3]:

S-ARP is based on digital signing of ARP packets by using public/private key pairs. Any S-ARP enabled host is identified by its own IP address and has public/private key pair. A simple certificate provides the binding between the host identity and its public key. Besides the host public key, the certificate contains the host IP address and the MAC address of the Authoritative Key Distributer (AKD), a trusted host acting as key repository. Each host sends its signed certificate containing the public key and the IP address to the AKD.

Every host maintains a ring of the public keys and the corresponding IP addresses previously requested to the AKD. When a host receives an S-ARP reply, it searches the sender IP address and its corresponding public key in its ring. If it finds such an entry, it uses the content to verify the signature; otherwise it sends a request to the AKD for the certificate. A request to the AKD is sent also in case the key in local ring does not verify the signature, since it may no longer be valid. The AKD sends a signed reply with the requested public key and the current time-stamp. Upon receiving the reply from the AKD, the host stores the public key in its ring and verifies the signature.

A drawback of this scheme is that the AKD constitutes a single point of failure in the network. If the AKD is down, a host can not verify ARP packets that are sent by a previously unknown host. Even if the AKD is working properly, an attacker can impersonate a host that goes down by cloning the MAC address of the host (but only until the cache entry of the host being impersonated, in the host being attacked expires). [4]

### 3.3 Intrusion Detection/Prevention Systems (IDS/IPS):

IDS/IPSs can be divided as host based and network based. Host based ones are installed on hosts and detect or protect for only those hosts. Network based ones listen to mirror port of the switch or some ports of the switch. They can detect or protect the hosts connected to those ports.

IDSs like Snort [10] are usually able to detect ARP attacks and inform the administrator with the generation of an appropriate alert or alarm. The main problem with IDSs is that they tend to generate a high number of false positives (alarms that turn out to be not part of attacks), that for them to be effective, it becomes a must for the company to put somebody in charge of dealing with these events. For many companies, having to pay somebody to do this job is not an option [4].

IPSs protect the host or network by sending TCP reset packet to the attacker or they can order switch to stop communication of the attacker.

### 3.4 Static ARP Entries:

As I explained in section 2.1 every host has a ARP cache in which <IP, MAC> pairs are kept for a limited time. If these entries are flushed after an aging time, these are called dynamic entries. On the other hand we can enter data to this cache manually. The data entered manually are called static entries and they have no aging time. An attacker can not change these entries by sending forged ARP replies. Static ARP entries are guaranteed solution for ARP spoofing. Although it is the guaranteed solution it is very difficult to implement. It is very hard to enter all the hosts <IP, MAC> pairs into all hosts in the network. If you entered once, maintenance of it also were very difficult. Because there always be change of IP addresses and some hosts are added or removed from the network. Furthermore, some operating systems (such as Windows) may accept ARP replies and update for static entries. [11]

This solution can be implemented for only limited number of hosts. For example <IP, MAC> pairs of intranet web servers or other important servers or <IP, MAC> pair of the network gateway can be entered manually to all hosts, thus ensuring the one way communication between them.

### 4. PROPOSED SOLUTION
### 4.1 Layer-2 Switch Operation and Filtering

LANs were based on bus topology at first. Later hubs began utilizing star topology. Hubs and bus networks create a common network segment which is shared by all hosts connected to that network. Common network segment creates one big collision domain which reduces network performance and allows a host to listen to other hosts communicating on the network. After hubs, switches were invented. Switch makes micro segmentation which means there is no collision and a host can not listen to other communicating pairs. At first switches were running at OSI layer-2, later layer-3,4 switches were invented.

A switch running at OSI layer-2 is used in this proposal. This switch has the ability to make filtering by looking at MAC addresses and protocol type of Ethernet frame. Normally switch works as follows: It looks at the destination MAC address of the incoming frame and tries to find it in its forwarding table. Switches keep <MAC, Port #> pairings in the forwarding table. If it finds the MAC address in the table then forwards the frame to the corresponding port. If it doesn't find the MAC address then it forwards the frame to all its ports except the one it is coming from. It also looks at the source address of the frame and if it is not in the table it enters the <MAC, Port #> pair in the forwarding table.

The switch used in the proposal looks additionally at the source or destination MAC address and protocol type of the frame and applies the rules entered by the administrator of the system. In the switch used here these rules state that no host on the network can answer the ARP requests except the ARPserver.

## 4.2 ARP Implementation of Operating Systems

**Windows (2000/XP) [12]:** ARP implementation of Windows 2000/XP obeys the rules of RFC 826 [1]. When a host receives an ARP request not asking for its own MAC address, it looks at the sender IP, if this IP is in the ARP cache then it updates the life time of this entry. If the ARP request is asking for its own MAC and the IP of sender is not in the cache then it enters IP and the MAC of sender to the cache. This means unsolicited ARP messages does not create an entry in the cache. If they exist, lifetime of this entry is updated.

**Solaris (9.0):** When Solaris 9.0 receives an ARP message either unsolicited or destined to itself it enters the sender IP and MAC pair to its cache if they are not in the cache. It keeps track of two lifetimes for these entries, one for unsolicited ARP entries and one for the others. Unsolicited ARP entries have 5 min. default lifetime and can be updated 3 times to a maximum of 15 min. The normal ARP entries have a lifetime of 20 min.

**Linux (Pardus:** A Linux distribution created in Turkey**):** ARP implementation of Linux also obeys the rules of RFC 826 [1].

## 4.3 MAC Spoofing With Ettercap [13]

Ettercap is a sniffer but it is also a powerful and flexible tool for man-in-the-middle attacks. It supports active and passive dissection of many protocols (even chiphered ones) and it includes many features for network and host analysis.

I started Ethereal (Now Wireshark) [14] on host with IP 192.0.0.108 to sniff packets for further analysis and from command line entered the following command.

>ettercap –i dev1 –hexview –Nad –H 192.0.0.1-2 192.0.0.15 192.0.0.100 00:0b:cd:b6:3e:a2 00:08:c7:9f:bd:a8 > den4

With this command my computer enters between the computers 192.0.0.15 and 192.0.0.100. Every packet exchanged between these computers is dumped into the file "den4". After Ettercap is started, I stopped Ethereal and inspect the packets sent from my computer by Ettercap.

| No. | Source | Destination | Protocol | Info |
| --- | --- | --- | --- | --- |
| 1 | 192.0.0.108 | Broadcast | ARP | Who has 192.0.0.100? Tel 0.0.0.0 |
| 2 | 192.0.0.100 | 192.0.0.108 | ARP | 192.0.0.100 is at 00:08:c7:9f:bd:a8 |
| 3 | 192.0.0.108 | Broadcast | ARP | Who has 192.0.0.15? Tel 0.0.0.0 |
| 4 | 192.0.0.15 | 192.0.0.108 | ARP | 192.0.0.15 is at 00:0b:cd:b6:3e:a2 |
| 5 | 192.0.0.100 | 192.0.0.15 | ICMP | Echo (Ping) request |
| 6 | 192.0.0.108 | 192.0.0.15 | ARP | 192.0.0.100 is at 00:0e:7f:5f:ba:40 |
| 7 | 192.0.0.15 | 192.0.0.100 | ICMP | Echo (Ping) request |
| 8 | 192.0.0.108 | 192.0.0.100 | ARP | 192.0.0.15 is at 00:0e:7f:5f:ba:40 |

Explanation of above packets is as follows:

With packet numbers 1,2,3,4 Ettercap learns the MAC addresses of victim hosts (which are 192.0.0.100 and 192.0.0.15). With packet 5 Ettercap sends an ICMP echo request by forging a packet with source IP 192.0.0.100 and destination IP 192.0.0.15. In this way an entry in the ARP cache of 192.0.0.15 machine is created. With packet 6 the entry created with packet 5 poisoned with the attacker MAC address. By packets 7 and 8 the same above is applied to 192.0.0.100 and this host also is poisoned with MAC address of attacker.

## 4.4 Design

### 4.4.1 Setup

In the proposed solution a layer-2 switch which has the capability of making decisions based on MAC address and protocol type constitutes the network infrastructure. All hosts are connected to this switch along with the host on which ARPserver application runs. Figure 6 shows the setup.

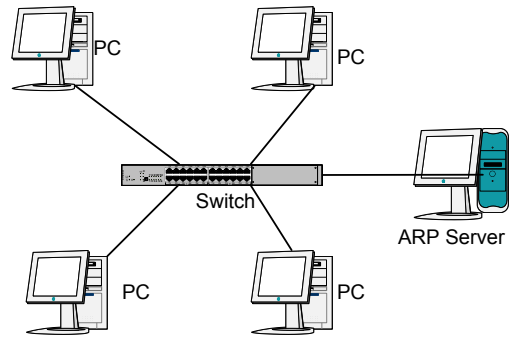

**Figure 6  Setup**

The following filtering rules are applied to all ports of the switch except the port to which ARPserver host is connected.

If (a Frame is inbound) and

    (Protocol_Type=0806) and

    (Source_MAC = MAC_of_ARP_server)

Then

    Drop Frame

If (a Frame is outbound) and

    (Protocol_Type=0806) and

    (Source_MAC <> MAC_of_ARP_server)

Then

    Drop Frame

The above pseudo code does the following:

All inbound frames (frames coming from the hosts to the switch) are checked. If the frame Ethernet protocol type is 0806 (ARP request/response frame) and source MAC address is the same as MAC address of the ARPserver host then the switch drops (does nothing about the frame) the frame. This rule guaranties that no host can spoof the MAC address of the ARPserver host.

All frames outbound (frames going from the switch to the hosts) are checked. If the protocol type is 0806 and source MAC address is not the same as MAC address of the ARPserver host then switch drops the frame. This rule guaranties that only ARPserver host can respond to ARP requests.

When these two rules are applied to switch, all ARP packets are destined to ARPserver host. Other hosts are prohibited from answering ARP request.

*4.4.2 ARPserver*

ARPserver application replies all ARP request packets broadcasted, besides checking for ARP attacks. It has a database of <IP, MAC> pairs. The problem is how to fill this database with all the active hosts' IP and MAC addresses.

**a)** The pairings can be learned by sending ICMP Echo request packets to all the hosts on the network in regular intervals. The length of interval may be determined by the rate of change of IP addresses and the rate of hosts attached or removed from the network.

**b)** ARPserver can continuously listen ARP broadcast packets and enter the sender IP and MAC addresses to its database.

**c)** Another version of ARPserver which has the capability of understanding DHCP packets may be implemented. If there is a DHCP server on the network, then it is better to install the ARPserver onto this host. During the initial IP address assigning process ARPserver learns <IP, MAC> pairs.

**d)** We can enter manually <IP, MAC> pairs to this database. These entries should be gateway and other important servers such as domain controller, DNS server e.g.

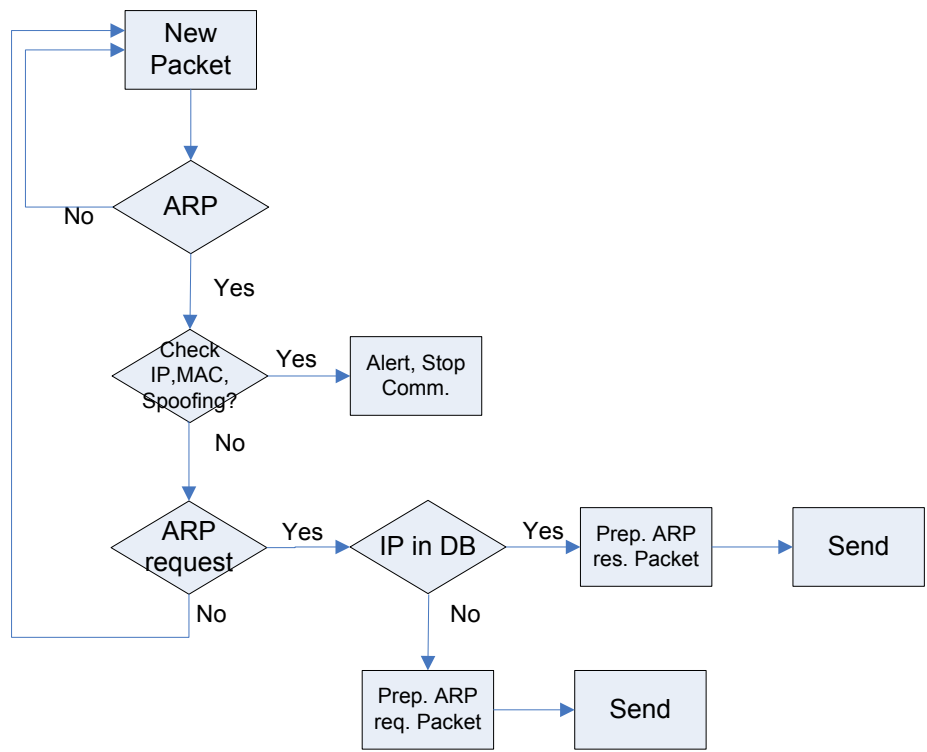

**Figure 7 ARPserver Flow**

ARPserver generally carries out the flow shown in Figure 7. The algorithm applied in "Check IP, MAC, Spoofing" decision box is as follows:

*If  (Ethernet_Source_Addr <> Sender_Ethernet_Addr)*
  *Then Alarm(Sender_Ethernet_Addr)*           // Sender is trying to hide itself
*Else*
  *Search <IP, MAC>_Table*
    *If (Sender_Ethernet_Addr in <IP, MAC>_Table & Sender_IP_Addr not equal to <IP$_{sender}$, MAC>_Pair*
      *Then change IP in <IP, MAC$_{sender}$>_Pair*           // A host may have two IP addresses

> If (Sender_IP_Addr in <IP, MAC>_Table & MAC in <$IP_{sender}$, MAC>_Pair !=
> Sender_Ethernet_Addr)
>  Then begin
>   Send_ARP_Request (
>    Ethernet_Dest_Addr:=MAC_of_<$IP_{sender}$, MAC>_Pair
>    Ethernet_Source_Addr:=$MAC_{ARPserver}$
>    Op:=Request
>    Sender_Ethernet_Addr:= $MAC_{ARPserver}$
>    Sender_IP_Addr:=$IP_{ARPserver}$
>    Target_Ethernet_Addr:= ?
>    Target_IP_Addr:= $IP_{sender}$
>    )
>   Wait_ARP_Reply
>   If ARP_Reply comes
>    Then Alarm(Sender_Ethernet_Addr) // Sender is tring to spoof another host
>   Else
>    Change MAC in <$IP_{sender}$, MAC> // Original host down and its IP is given to another host
> 
> If (Sender_Ethernet_Addr not in <IP, MAC>_Table)
>  Then enter ($IP_{sender}$, $MAC_{sender}$) into Table

## 4.5 Implementation

### 4.5.1 Switch and ACLs:

For the implementation and tests of the proposal a Cisco 2960 [15] gigabit switch is used. Setup shown in Figure 8 is prepared.

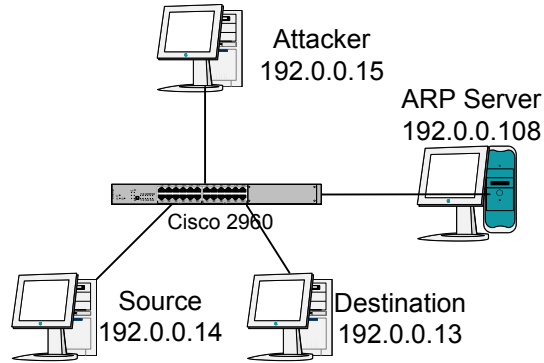

**Figure 8 Implementation Setup**

Unfortunately Cisco 2960 does not support ACLs to be applied to outgoing packets. This is in mind following ACLs are created and applied to switch ports:

Switch> enable

Switch# configure terminal

Switch (config)# mac access-list extended ARPblok

Switch(config-ext-nacl)# permit any host 000e.7f5f.ba40 0x806 0x0

Switch(config-ext-nacl)# permit any host ffff.ffff.ffff 0x806 0x0

Switch(config-ext-nacl)# deny any any 0x806 0x0

Switch(config-ext-nacl)# permit any any

Switch(config-ext-nacl)# end

Switch(config)# interface gigabitethernet0/19

Switch(config-if)# mac access-group ARPblok in

Switch(config-if)# exit

Switch(config)# interface gigabitethernet0/20

Switch(config-if)# mac access-group ARPblok in

Switch(config-if)# exit

Switch(config)# interface gigabitethernet0/21

Switch(config-if)# mac access-group ARPblok in

Switch(config-if)# exit

*4.5.2  Server Software:*
ARPserver software developed here is not a comprehensive one. The aim is just to show that it is capable of preventing ARP spoofing attacks.

ARPserver software will be developed using open source packet capture and filter library called WinPcap [16]. When WinPcap [16] is installed on a computer, packet.dll and npf.sys files will be loaded to computer. These files communicate with the Ethernet driver. We communicate to these files using packet.api. The software was coded using Delphi 7.0 and the interface units for WinPcap [16], originally written by Lars Peter Christiansen. In ARPserver software there is a simple sniffer which utilizes this API. Sniffer has an event handler which will inform about that a packet has arrived. Sniffer will examine the packet. If it understands that it is an ARP packet then it does the flow shown in Figure 7. The software keeps two databases in memory, one for <IP, MAC> pairings and the other for spoofed MAC address list. <IP, MAC> pairings database keeps at most 254 entries which is the maximum limit for a C-class network. Before updating the database all entries in the database are checked to see if the new entry is a spoofed one. This sniffer has also the capability of preparing and sending ARP response packets. ARPserver responds all ARP requests except for the ones asking the ARPserver host MAC address.

**4.6  Tests**

*4.6.1  Connectivity Test:*
ICMP echo request packets are sent before starting ARPserver software from 192.0.0.13 to 192.0.0.14 and 192.0.0.15 and no responses come back. After starting ARPserver connectivity is established and ICMP echo responses come.

*4.6.2  IP Change:*
IP of 192.0.0.13 is changed to 192.0.0.17 which is not present in the network. No problem is observed. Successfully connectivity is established with other hosts.

*4.6.3  IP Conflict:*
Host IP of 192.0.0.13 is changed to 192.0.0.14. This situation can be thought as also MAC spoofing, because 192.0.0.13 is announcing its MAC as the MAC address of the host with the IP 192.0.0.14. ARPserver puts the host originally 192.0.0.13 into spoofing list.

*4.6.4  MAC change:*
One can change the MAC address of his/her computer using software such as SMAC [17]. By simply changing the MAC address it is almost impossible to access confidential data. Switches do not allow two same MAC addresses to be exit in their <MAC, port#> table. Because every <MAC, port#> pair has an minimum aging time, a malicious host changing its MAC to a communicating Hosts' MAC address can not access the switch. This can be used as DOS (Denial of Service) attack. ARPserver can understand this from rapidly changing IP of the <IP, MAC> pair. Actually port security is the way of preventing this.

*4.6.5 IP and MAC change:*
If someone changes his/her computer's IP and MAC to someone else's, then ARPserver can't recognize this. The prevention method of this is also port security.

*4.6.6 Using Ettercap [12]:*
Ettercap [12] can not change the ARP caches of the hosts. It is obvious from the ACLs that sending direct ARP packets are blocked by the ACLs, which is the way of Ettercap [12].

### 4.7 Future Work

a. ARPserver performance will be tested. Tests will be made on heavy load to see if it is causing any connection loss or denial of service.

b. ARPserver is single point of failure. If somehow does not work then whole network stop communicating. So another ARPserver host will be connected to network as redundant or working as load balancing.

c. This software can be expanded as to run with mirror ports where MAC based ACL are not available.

d. A small client software which will be loaded to all hosts in the network and forward all unicast ARP packets to ARPserver can be implemented for use where it is not possible to utilize ACLs.

## 5. CONCLUSION

This paper tried to explain ARP spoofing and proposed a solution to it and finally implementation was explained. There had been several proposals in the past as it is explained in section 3 of this document and as in [4]. All proposals have some advantages and disadvantages. But for the time being there is no widely available and known solution to this problem.

Companies spend lots of money for firewalls and virus protection software. But they don't know how dangerous internal threats can be. A malicious worker can easily obtain passwords and the company confidential data through ARP spoofing. The ARPserver software like the one explained here will be very helpful to network administrators to defend their network against ARP spoofing.